# Beyond the Ultra-Dense Barrier -
## Paradigm shifts on the road beyond 1000x wireless capacity

Jens Zander

*Wireless@KTH, KTH – Royal Institute of Technology, Stockholm, Sweden

**Abstract**

It has become increasingly clear that the current design paradigm for mobile broadband systems is not a scalable and economically feasible way to solve the expected future "capacity crunch", in particular in indoor locations with large user densities. "Moore's law", e.g. state-of-the art signal processing and advanced antenna techniques now being researched, as well as more millimeter wave spectrum indeed provide more capacity, but are not the answer to the 3-4 orders of magnitude more capacity at today's cost, that is needed. We argue that solving the engineering problem of providing high data rates alone is not sufficient. Instead we need to solve the techno-economic problem to find both business models and scalable technical solutions that provide extreme area capacity for a given cost and energy consumption. In this paper we will show that achieving very high capacities is indeed feasible in indoor environments. However, to become economically viable, approaches with radically different fundamental cost factors compared to those used in today's cellular systems are needed. To reach very high capacity we must venture beyond *the ultra-dense barrier*, i.e. networks where the number of access points in an area is (considerably) larger than the active number mobile terminals. In such networks area capacities of more than 1 Gbit/s/m$^2$ are perfectly feasible. The problem set encountered in such Ultra-Dense Networks (UDN) is very different from conventional cellular systems and their solution requires conceptually new tools. We will address some of the fundamental aspects and performance limits, modeling of propagation, deployment and user traffic, and discuss the techno-economics of various network architectures. Finally we will summarize some of the most significant unsolved research questions in the field.

# 1. Introduction

The success of mobile and wireless access to the Internet has been nothing short of monumental. LTE or "4G", being the "dominant design" for wireless wide area mobility, has rapidly become a worldwide commercial success. "Always connected" with high-date rate access (almost) everywhere has sparked an evolution where "apps" in smartphones provide a plethora of cloud-based mobile services, literary in anyone's pocket. As with previous generations, time has now come for the mobile (infrastructure) manufacturers to look ahead to research the next generation of wireless technology, which naturally has been labeled "5G". Similar to the situation after 2G and 3G, the industry is now looking for, if not the "killer app"(although there is not likely to be one this time either), then at least some new service scenarios that cannot be managed with existing technology and thus would create a demand for some new, revolutionary technology. The European Project METIS-2020 is one example of a systematic search for these new requirements and the technologies to meet them[1]. There now seems to emerge an industry consensus, that the requirements for 5G and beyond roughly boil down to two main areas: *Containing the "Data Tsunami"*, i.e. to provide (much) "more of the same", i.e. more capacity and higher data rates to quench the thirst of more users for more "data" and catering for *efficient Machine-Type Communication*(MTC)" - systems allowing both billions on "things" to communicate as well as providing high-reliability, low-delay wireless communication in real-time control-loops.

In this paper we will mainly address the first challenge. It is a historical fact that the wireless data traffic has roughly doubled every year over the last 7-8 years in most markets and is expected to keep growing in a similar fashion in the years to come. The main underlying driver for this increase is that the internet paradigm, i.e. Internet Protocol (IP) based access has become the "dominant design" also in wireless communication [3]. The internet paradigm has taken over completely, not because of its technical superiority but due to its high degree of flexibility, providing a "future-proof" platform for all kinds of applications. The price we pay for this flexibility is inefficiency in the transmission medium due to the large overhead caused by many protocol layers that in turn will increase the need for capacity. One may say that we trade network capacity for higher service transparency. In the world of clouds and the internet, however, the marginal cost of a transmitted bit should be zero. In practical wireless systems, however, providing this capacity is not entirely for free. Extrapolating the above trend has given rise to the challenge "1000 times more (area) capacity at today's cost and energy consumption"

In[3], technical and architectural solutions that have a realistic possibility to achieve these targets have been explored. It has become obvious that "Moore's law", i.e. even more powerful signal processing will not again "save the day" in high-speed wireless access. In [3], we further argued that additional improvements of the PHY-layer are possible, but it is not likely that these alone will provide a solution to the capacity problem. The increases in peak user data rate new PHY-layer technologies seem to provide are of course helpful, but these rates have to be shared between all users in the same area. The way to increase capacity by orders of magnitude in wireless *networks* is, and has always been, to *densify the infrastructure*, i.e. to increase the number of base stations per area unit. A key observation is that we are no longer facing only an engineering problem – we are dealing with a *techno-economic problem*. The problem is to provide the high access data rate under given resources constraints. Some of the sub-problems involved are technical (e.g. available bandwidth, energy consumption, noise), but most of them are economical (e.g. number and cost of base stations, wired infrastructure). It is a well-known result fact that the traditional cellular concept does not scale well, as the required number of base stations (and thus the cost) grows roughly proportional to the capacity required[10]. This worked well in the era of mobile telephony as more capacity meant more paying customers – now the same users expect much higher data rates without paying more. It is clear that devising a new network architecture is necessary to break this vicious cost circle [4].

There is hope, however, in the fact that the exponentially growing traffic is not uniformly distributed. The evolution of *heterogeneous cellular networks ("HetNet")* with a mix of "micro" and even "femto" cells with the aim to match the traffic demand in various locations[5][6]. Most traffic is found in dense population centers and mainly indoor. This case is described in the "Amazingly Fast" scenario of the 5GPPP METIS project, or in the "Pervasive video" scenario of the the NGNM or the "Gigabit in a second" scenario by the ITU-R (summarized in[2]). Recent studies suggest that it might be possible to meet the increase in capacity demand by super-dense deployments, provided this deployment can be made at a very low cost. When cell sizes keep decreasing, the traditional mobile operators literary run into a "brick wall". The "brute force" solution to reach users inside the buildings by penetrating walls and windows with high power transmissions is neither very reliable, nor very efficient. In order to increase the base station density further we need to deploy them indoor. Inside the walls, however, a radically different techno-economic landscape opens up in which very different rules apply. The traditional operators are neither able to use their usual toolbox for deployment, nor do they have the business models for indoor operation. These are the challenges addressed in this paper.

## 2. What makes the indoor environment so very different ?

In the indoor environment walls, metalized windows etc., create high propagation losses for the signals from the access points (AP:s) to the user terminals. In usual cellular settings, this would be a significant problem, labeled as "poor coverage". However, in very high-density systems, where there is an access point ("base station") in almost every room providing almost line-of-sight wireless links, coverage is rarely a problem. From an interference perspective, the wall attenuation becomes a "blessing", as interference from other access points in different rooms, may be effectively blocked – in particular in centimeter and millimeter wave systems (above 5 GHz). As very little interference escapes from a building, the same spectrum can be reused in adjacent buildings [7][10]. In fact, this phenomenon also opens for alternative ways to access more frequency spectrum. Indoor low-power systems operating above 6 GHz may successfully share vast amounts of spectrum with other outdoor, wide-area services on a secondary basis [12]. Large amounts of available spectrum can be used to further improve the performance as well as lowering the equipment cost and energy consumption.

| Characteristic | Cellular, Wide-area paradigm | High-Density, Short Range |
|---|---|---|
| Propagation | Distance loss, shadowing, rich multipath | Mostly LOS, Body shadowing |
| Interference | Interference sum of many components (averaging) | Extremely varying interference |
| Duplexing | Up & Downlink have different characteristics (power) and must be separated | Link direction irrelevant Access points/terminals can be defined at higher layers. |
| Engineering limitations | Range, Interference, Energy | Interference |
| Peak rate limitation set by | Noise & Interference | Equipment (very high SNR) |
| Cost limitations | Sites: Acquisition, Antennas, Equipment, Deployment, Backhaul, Spectrum licenses | Backhaul, Deployment |
| Active Users/Base station | 1-100 | 0,01 - 1 |
| Available radio bandwidth | < 0,5 GHz Licensed | > 5 GHz Secondary sharing |
| Business model | Subscription based service Per month or per MB charging | Free to all tenants and visitors in building (similar to A/C, lighting, running warm water) |
| Design paradigm | Industrial grade, Centralized control, "mandatory complexity" | Consumer grade, Distributed control, plug-and-play |
| Maintenance model | Single point of failure - 24/7 monitoring | Graceful degradation – replace when time available |

Table 1: Comparison of traditional cellular and high-density system characteristic

There are also large differences regarding the techno-economics of indoor systems in comparison with the outdoor, cellular variety. Equipment cost for indoor operation is significantly lower and the cost of maintenance is only a fraction of the outdoor counterpart. In high-density "plug-and-play" deployments, the failure of a single access point may not even be notable by the users, as it results only in a moderate loss in performance. The access point may be replaced at some convenient time. This means that a network built with consumer grade access points can provide high reliability, due to the high degree of redundancy. Access point hardware cost is likely to be in the 100's of USD, rather than in the 1000's of USD expected for outdoor deployment. Looking at the total cost for a base station/access point (the "site cost"), the difference is even more striking [9]. Outdoor system costs are dominated by various physical items like towers, antennas, buildings, energy, backhaul connections etc. All these cost items literary vanish for low power indoor access systems. What remains is the cost for the wired infrastructure, the *backhaul* for the access points. As the number of access points becomes very large in high-density systems, the backhauling cost becomes the dominant cost factor. Since the density of access points is very high, only a very few users with high rate demands will share the cost of each access site and the associated backhaul. This means that, despite the low cost of equipment, the cost per user still may be significant in indoor systems.

A consequence is that the usual, competitive, public operation business model used in mobile cellular collapses. The operators cannot afford each of them separately "wiring" a certain building or home to deploy their own access points. The only economically sensible thing is that the facility/home owner deploys his own network, which then could be shared by the public operators for serving their indoor customers.

This is indeed nothing new - in fact this is already the dominant business model for wired/fixed communication. Public operators connect the building, but the internal wired network is provided by the facility owners themselves. Most office buildings are already fitted with Cat-5/6 Ethernet cabling that provides IP-connectivity in almost every room. Any wireless access technology that can use such an existing infrastructure will have a monumental cost advantage compared to the traditional operator model. To keep deployment costs at bay, concepts and technologies that allow for low cost deployment, distributed solutions and "plug and play" are of paramount importance. The user experience when deploying a WiFi-network can here be seen as role model (low cost, "out of the box", self-configuring).

The perception of the frequency spectrum is another dividing line between the two paradigms. Traditional mobile operators build their business on having exclusive access to nation-wide blocks of spectrum. In order to create a world market for mobile terminals, these frequency blocks are harmonized over large parts of the globe. Needless to say the spectrum that fulfills all these requirements is scarce. In addition, nation-wide block of spectrum gives the operator the opportunity to services millions of potential users ("pops"). This means that the competition for this spectrum will be fierce and large sums are paid at spectrum auctions for "cellular spectrum". A further consequence of the spectrum scarcity is the massive complexity of the engineering solutions required to squeeze 100's of Mbit/s into narrow radio channels. The result of spectrum scarcity has been (and will be) high equipment cost and energy hungry solutions.

The spectrum used by a facility owner for his indoor system at 5 GHz and above, is a completely different creature. The signals hardly leave or enter the building and the spectrum can be reused next door. There should be no competition for the spectrum as the facility owner is the only one that can make use of it. Finally, the number of users in the building ("pop") is a few hundred. A hypothetical auction for the indoor spectrum in a specific building would therefore result in a zero spectrum price. In addition, as we mentioned previously, most of the spectrum above 5 GHz would probably lend itself quite well for secondary reuse by indoor, low-power devices[12]. As an alternative, we can go into the mm-wave range where there are still large chunks of spectrum available or even into the near-visible light range ("Li-Fi")[13]. Low cm-wave spectrum (<10 GHz) has several advantages: equipment is thus likely to become available at low cost within a short time span and it allows user devices to work even when it is worn in a pocket or briefcase. The disadvantage is the resistance from incumbent users of this spectrum. mm-wave and (near) visible light systems on the other hand have the advantage of large chunks of spectrum being still available, they provide even better isolation from out-of-building interference and they allow for the use of compact, high gain directional antennas that can improve the signal-to-interference ratio [7][13]. Their disadvantages are mainly the susceptibility to body-shadowing. Although extensive prototyping is currently in progress, the device technologies are still relatively unproven in the mass market.

In either case a reasonable licensing regime would be to give the facility owner access to <u>all</u> spectrum, on the condition that public operators are given the opportunity to use the indoor network on "fair and reasonable" terms, e.g. if the facility owner provides WiFi-like access to all it's tenants, all the (outdoor) mobile operators should have the opportunity to off-load their indoor traffic to this network.

Moving from the "per-minute charge"-paradigm to a "flat rate" per –month charging model is radically changing the business of the operators. In the old paradigm, "Handing-off" ("off-loading") a user to some other network meant loosing call-minutes and revenues. In the flat-rate paradigm, keeping the customer's business while letting someone else provide the network access, makes "off-loading" an interesting business proposition[6]. There is now ample proof of that virtually seamless wireless IP-access for slowly moving users doesn't require a single, monolithic system. Smartphones today automatically switch to the users private WiFi network when he comes home. Many cellular operators now provide WiFi "offloading" schemes also in public "hot-spot" environments. The mobile operator's business is transitioning from solely providing infrastructure for communication to manage the connectivity of their customer.

Table 1 summarizes the key differences in characteristics between the two paradigms. By now, it should be clear, that the engineering and business rules in the two "worlds" differ significantly and that it is not likely that indoor solutions based solely on the traditional cellular paradigm will be successful. In the next section we will discuss some of the characteristics a dense wireless network should have to provide capacities in excess of 1 Gbit/s/m$^2$.

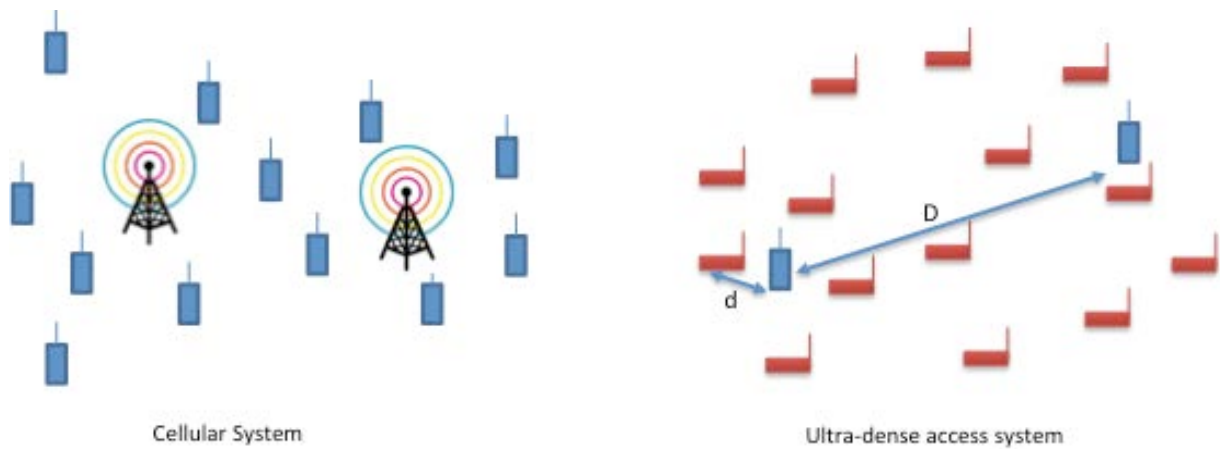

Fig 1. Conventional cellular system (many terminals per base station) vs Ultra-Dense Access System (many access points per access point).

## 3. Beyond the "Ultra-Dense barrier"

The term *ultra-dense network* (UDN) has been frequently used in recent scholarly work. Most of this work actually refers to dense cellular networks, typically with inter-base station distances (IBD) from a few ten's to a few hundred meters [5]. These are, in fact, base station densities already found in many urban micro/pico-cells where most of the "cellular wisdom" still applies. To reach extreme capacities we need to study systems with even higher access point densities, corresponding to IBDs in the order of meters. As we note in the previous section, the absolute distance to the access point will affect the propagation conditions. These become more of "line-of-sight" character, which means the potentially achievable data rates increases rapidly. Although the benefits from some of the most popular statistical signal processing techniques (e.g. MIMO) are diminished, transmission rates of 10's of Gbit/s should still be feasible over the short ranges. With a "cell size" of less than 10m$^2$, we would reach our capacity target of 1 Gbit/s/m$^2$. The key problem in such a system is how to deal with the almost equally rapidly increasing line-of-sight interference from other access points and terminals. To assess the impact of interference, the *relative density* of access points is an important parameter. If we aim at serving a fixed user population with a large traffic demand and keep increasing the density of access points, at some point there will be more access points than terminals. We denote this point the "*ultra-dense barrier*" or, more precisely, where the access point density exceeds the user density.

As we push far beyond this point, the character of the system changes radically, and the behavior becomes similar to a distributed system of antennas (fig 1). In its simplest form each terminal will connect to the nearest access point which is only a few meters away (*d* in fig 1). The wanted signal strength will be extremely high, but so will the potential interference from other access points and terminals. Since there are few terminals, only the serving access points need to be switched on whereas all the unused access points can be kept "silent".

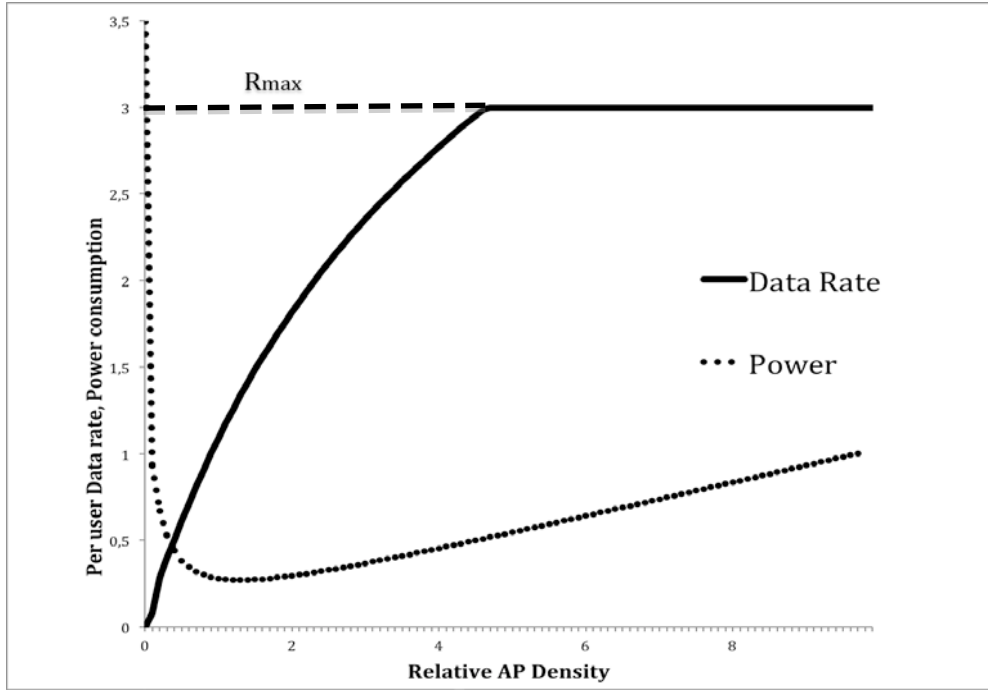

Fig 2 Illustration of eqs 1 & 2: Data rate and power consumption (per user) as function of the relative AP density $\lambda_{AP}/\lambda_U$ in ultradense wireless Networks

This means that the interference is mainly caused by the nearest active user at distance *D*, and not by the nearest access point that is not serving our targeted user. As a consequence, the relative reuse distance *D/d* that determines the Signal-to-Interference ratio, may still be quite large.

Using Stochastic Point Processes to model AP and terminal locations in dense networks has become a popular technique to study dense wireless networks. This technique allows us to assess the asymptotic behavior of dense networks. Despite the growing interference, the analysis in [12] shows that area capacity keeps growing as we increase the user density, even in (near) free-space environments. Assuming the user and AP densities are $\lambda_U$ (users/area unit) $\lambda_{AP}$ (access points/area unit), that propagation loss follows a power law with constant $\alpha$, we can simplify the results in [12] and write the area capacity (bit/s/m²) as [1]

$$\text{Area capacity} \propto \begin{cases} \lambda_U W_{SYS} \log\left(1 + c\left(\dfrac{\lambda_{AP}}{\lambda_U}\right)^{\alpha/2}\right) & \lambda_{AP} \leq \lambda_{AP}^*(R_{\max}) \\ R_{\max}\lambda_U & \lambda_{AP} > \lambda_{AP}^*(R_{\max}) \end{cases} \qquad (1)$$

---

[1] With received power inversely proportional to the distance to the power $\alpha$, the Signal-to-Interference Ratio in this interference dominated scenario becomes proportional to $(D/d)^\alpha$ if we use the notation i figure 1. As D>>d in an ultradense deployment, AP:s and users are close, which means that the distance to the nearest user and his associated AP is the approximately the same. The expected distance *D* and *d* are proportional to $1/\sqrt{\lambda_{AP}}$ and $1/\sqrt{\lambda_U}$ respectively and we can use the Shannon formula to find the approximate expression above.

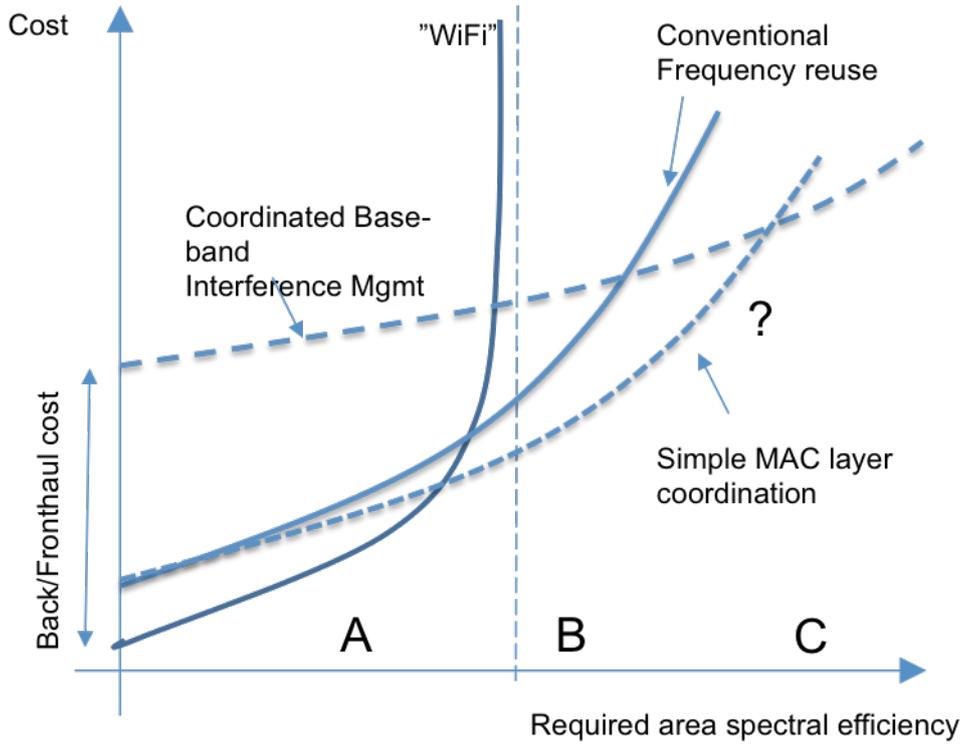

Fig 3 Cost for providing a certain area spectral efficiency (adapted from [10] )

where $W_{SYS}$ is the available (spectrum) bandwidth, $R_{max}$ is the peak data rate of the system and $c$ is a constant that includes the combined effect of interference suppression measures (interference cancellation, beam-forming etc).

The expression clearly shows that, for a given user density, the area capacity increases monotonously with the access point density until we get limited by the peak data rate of the equipment. Beyond this point it does not pay off to further increase the AP density as, even though the SIR keeps increasing, every user is already served the peak data rate.

An important concern related to the scalability is the energy consumption in very dense networks. Using the same model as above it is easy to see that the system energy consumption per user) [15] can be written as

$$\text{Power/User} = \frac{c_1}{\lambda_{AP}^{\alpha/2}} + c_2 \frac{\lambda_{AP}}{\lambda_U} \qquad (2)$$

where the first term corresponds to the transmit power that approaches zero with increasing $\lambda_{AP}$ as the nearest AP comes closer and closer to the user. The second term corresponds to the idle power consumption of the AP:s that are currently not used. This term will keep growing proportionally to the AP density. Fig 2 illustrates the average data rate per user (the Area capacity divided by $\lambda_U$) and power consumption per user as function of the relative AP density.

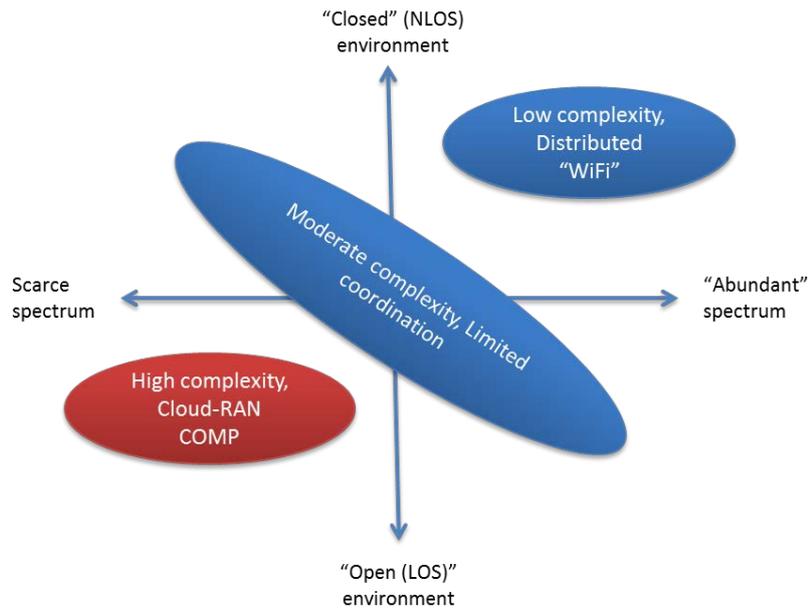

Fig 4 Architectural options for Ultra Dense systems

Note that, even with a limited peak data rate, there is no (theoretical) limit to the area capacity as the number of users ($\lambda_U$) increases, as long as we keep operating beyond the "ultra-dense barrier", i.e. the ratio $\lambda_{AP}/\lambda_U$ is kept large. Although this is an interesting theoretical result, a concern is that the rate only increases with the logarithm of the access point density and beam-forming gain c.

## 4. Architectural considerations

In the previous section we have shown that UDNs technically are capable of achieving any capacity by just increasing the AP density even in (near) LOS environments. In this section we will discuss some of the techno-economic limitations and how they may affect future system designs. As was shown in the previous section, the area capacity is (to some limit) an increasing function of the number of access points. Thus to achieve a certain capacity, a certain number of AP:s per area unit is needed. The cost to provide this capacity consists of the cost of the AP:s and the cost of the backhaul network. Fig 3 sketches the relative cost trends for some of the basic system design principles as function of the required area spectral density (area capacity divided by $W_{SYS}$) (adapted from [10]). We see two distinct grouping of systems:

- Systems with more or less distributed control functions. These systems can be deployed using existing, moderate performance backhaul (typically Cat 5/6 twisted pair cables)

- Centralized solutions (a.k.a. "Cloud RAN") with centralized base-band signal processing allowing for waveform-level interference coordination (e.g. "Coordinated Multipoint", COMP") These solutions provide very high capacities but require very complex processing and dedicated, very high speed, low latency "front-haul[2]" networks.

---

[2] "Fronthaul" denotes the wired (fiber) connection between the centralized signal processing facility and the base station, a.k.a. "remote radio heads".

In the first category we find WiFi-type systems and the traditional frequency reuse type "pico-cellular" systems (solid lines). The WiFi-type systems have a very low fixed cost as they use existing backhaul and do not need centralized control. Systems with advanced coordinated interference (dashed line) exhibit a high fixed cost for the front/backhaul, but can achieve high capacities at low cost (a high value of c in eq 1). WiFi systems, due to their "listen-before-talk" regime using a fixed power threshold and only very limited interference resilience, reach a capacity limit that cannot be overcome by adding more AP:s. Pico-cellular systems will on the other hand have a more moderately increasing cost. Eventually this cost will also grow faster and faster as the capacity is only increasing with the logarithm of the AP density in eq. 1.

The exact shape of the curves in fig 3 and which design principle is to be preferred will depend on the *effective amount of available spectrum* for our indoor system. If large amounts of spectrum (many GHz of bandwidth) would be available, the required area spectrum efficiency will be low (region "A") and low complexity, "WiFi-type" systems will dominate the scene. If there is a shortage of spectrum we are likely to operate in region B or C where centralized system designs with more advanced interference mitigation methods are needed.

The *intended usage scenarios* will also affect the effective amount of spectrum available. "Closed", walled-in environments like offices will at cm-wave frequencies and above basically limit the interference to those users and AP:s co-located in the same room. In "Open" environments like large shopping mall, railway stations and stadiums, we are likely to experience significant LOS interference from access points that are far away. "Closed" environments offer thus better opportunities for spectrum reuse and secondary spectrum sharing and thus effectively make more spectrum available than in corresponding "Open" our outdoor environments. Looking at the design options and their techno-economic feasibility, we map potential designs on the various scenario characteristics in fig 4. The upper right corner corresponds to an office or home scenario with "abundant" availability of spectrum in relation to the targeted capacity.

We are in "region A" of fig 4 with plenty of spectrum available, and there is no need for advanced interference management. The focus will be on low-cost, low-complexity equipment working on existing backhauls. As capacity demands today are mostly rather moderate, WiFi is already outcompeting other (cellular) technologies in this domain by a wide margin – not because WiFi has higher performance, but because WiFi is a technology that is able to exploit the large bandwidth offered in unlicensed spectrum, it offers low-cost plug-and-play deployment and simple co-existence with neighboring networks.

The left side of fig.4 corresponds to a scenario, where we have not been able to secure enough bandwidth and interference becomes a significant problem and low-complexity listen-before-talk schemes as WiFi run into severe problems. The required area spectrum efficiency is high and we enter region B or even C.

| | Technology | Area | # AP:s | $\lambda_{AP}/\lambda_U$ | Inter AP distance | Required spectrum |
|---|---|---|---|---|---|---|
| Small conf.room | **WiFi** | **20m²** | **3** | **0,75** | **2,5 m** | **480 MHz** |
| Cafeteria, Auditorium | **WiFi** | **150m²** | **21** | **0,75** | **2,5 m** | **3.3 GHz** |
| Cafeteria, Auditorium | **UDN** | **150m²** | **150** | **5** | **1 m** | **2 GHz** |
| Cafeteria, Auditorium | **UDN-BF** | **150m²** | **150** | **5** | **1 m** | **500 MHz** |

Table 2. Required spectrum to reach an area capacity of 1 Gbit/s/m² in an open area using current WiFi (IEEE 802.11ac). Assumptions: $\lambda_U$=0.2 users/m² WiFi: No frequency reuse, 7 Gbit/s peak data rate in 160 MHz channel. Hypothetical UDN: No beamforming (c=1), UDN-BF beamforming (c=20dB). Spectral efficiency is assumed to be the same in all scenarios.

In more closed environments (upper left quadrant) we could rely on pico-cellular frequency reuse (region "B"), whereas in open environments with limited spectrum we are back in a "cellular regime" where more active interference management is needed to improve capacity (region C).

In which region would we be if today's of-the-shelf technology would be used? Is it feasble to reach a capacity of more than 1 Gbit/m² at all ? Let us make a rough estimate of the orders-of-magnitude involved. The results are summarized in Table 2 that shows the estimated amount of spectrum needed to achieve 1 Gbit/m² in some simple scenarios using state-of-the-art (2015) WiFi technology (IEEE 802.11ac). Each 802.11ac access point is capable of providing a peak rate close to 7 Gbit/s in a separate 160 MHz channel. In the medium sized conference room (where the same spectrum can be reused "next door"), we are already there, since "only" 480MHz are needed which would fit in the current 5 GHz band. We are clearly in "Region A" where it will be hard to compete with WiFi-technology. In are larger open area,. such as Cafeteria or larger Auditorium the limitations of the WiFi system becomes obvious. Adding more access points does not help as the system is not capable of exploting spectrum reuse. Unless we can make several GHz of spectrum available, we move into "Region B" of fig 3 and some type of "pico-cellular" concept as discussed in section 4 has to be used. Increasing the AP density and employing some kind of interference management (here beamforming) again will make us acheive the required capacity at reasonable spectrum bandwidths. A challenge in this scenario is if the cost for such "smarter" access points can be kept low enough to allow massive deployment.

## 5. Discussion & Research challenges

We have demonstrated that UDN:s represent a feasible way to reach the 1000x capacity target in indoor environments – by a wide margin. From a strict engineering perspective, there are no limitations to the capacity that can be achieved, and area capacities in excess of 1 Gbit/s are thus clearly within reach. The question is instead if this can be done in scalable and affordable way. The key stumbeling blocks on the road to massive amounts of access points are related ot the *system complexity* and the *cost of the backhaul*. The answer to both questions seems to be found in solutions with highly distributed processing. This leads us to identify a few important items for future research:

- *Seconday spectrum availability*: Secondary use of large amounts of cm-wave spectrum would put us in the upper rigth quandrant of fig 4, where low-complexity WiFi solutions would be the answer to most problems. What are the co-existence criteria that would allow spectrum can be used in this way ? What is the protection provided by buildings at these frequencies? Would "free/unlicenced" indoor use of frequencies above 6 GHz be feasible?

- *UDN - complexity and backhaul cost*: Are there interference management solutions involving more modest, MAC-layer coordination, that can be implemented over legacy backhaul and that could the preferred choice over a wide range and capacity demands (dotted line in fig 3). If listen-before-talk does not seem to work in UDN:s, what MAC-layer coordination should be used instead ? Adaptive beamforming seems to be the most promising technique.

- *UDN – "Cloud RAN":* An interesting research topic and potential business proposition is how to manage the quality and performance in network with large numbers of "plug-and-play" deployed access points. Is centralized management of such moderately coordinated networks feasible over standard IP based connections with high capacity but very limited delay guaranties ?

- *UDN - energy management*: Can we use the above techniques to send non-used APs to "sleep" and quickly waking them up, together with effective "sleep modes" are essential features to keep the energy consumption in check.